\documentclass{elsarticle}
\usepackage[utf8]{inputenc}
\usepackage{setspace}
\usepackage{graphicx}
\usepackage{amsmath}
\usepackage[margin=1in]{geometry}
\usepackage{subfig}
\usepackage{float}
\usepackage{booktabs}
\usepackage[export]{adjustbox}
\usepackage{nomencl}
\usepackage{url}
\usepackage{framed} 
\newcommand{\bigb}[1]{\left( #1 \right)}         
\newcommand{\myb}[1]{\mbox{\boldmath $#1$}}      
\def\dotp{\mbox{\tiny $\bullet \ $}}             

\begin{document}

\doublespacing
\begin{frontmatter}


\title{Database development and exploration of process--microstructure relationships using variational autoencoders}
\author{Srihari Sundar\corref{cor1}}
\author{Veera Sundararaghavan\corref{cor1}\fnref{label2}}
\fntext[label2]{Corresponding author}
\ead{veeras@umich.edu}
\ead[url]{http://umich.edu/~veeras}
\cortext[cor1]{Department of Aerospace Engineering,  University of Michigan, Ann Arbor, MI 48105, USA}

\begin{abstract}
The paper demonstrates graphical representation of a large database containing process--microstructure relationships using an unsupervised machine learning algorithm. Correlating microstructural features to processing is an essential first step to answer the difficult problem of process sequence design. In this paper, a large database of 346,200 orientation distribution functions resulting from a variety of process sequences is constructed, where each sequence comprises up to four stages of tension, compression and rolling along different directions in various permutations. This open-source database is constructed for collaborative development of process design algorithms. The paper demonstrates a novel application of the large database: graphical representation of texture--process relationships. A variational autoencoder is used to reduce the entire database to a two dimensional latent space where variations in processes and properties can be visualized. Using proximity analysis in this latent space, we can quickly unearth multiple process solutions to the problem of texture or property design. 

\end{abstract}


\end{frontmatter}

\section{Introduction}
Properties of metallic alloys such as elastic modulus, strength, and thermal conductivity are dependent on the substructure of the material at lengths of few hundred microns and below. This microstructure is composed of an aggregate of micro--scale grains (or crystals) of various shapes and sizes. Each grain is distinguished by its crystallographic orientation, which quantifies spatial orientation of the atomic lattice.  Descriptors such as the probability density function for orientations: the orientation distribution function (ODF) describes the texture or the distribution of orientations in the microstructure and is an important determiner of properties in polycrystalline alloys. Given the texture, engineering properties can be computed through averaging relationships. The inverse problem of identifying microstructural features that lead to a given set of properties has been addressed in the past by several authors using optimization techniques \cite{mssl52,mssl68} or data-mining methods \cite{cecen2018material, jung2019efficient, decost2017exploring, fullwood2010microstructure, xu2014descriptor, mozaffar2019deep}. In the case of identifying textures that lead to a known set of properties, typically linear programming or gradient optimization methods \cite{mssl52,mssl68,fullwood2010microstructure} have been employed, while in recent times, data mining methods has been used for increased efficiency \cite{mssl29}. While these methods can identify optimal microstructures, the problem of designing processes to achieve optimal microstructures has comparatively seen less attention. 

Microstructures can be tailored so that desired properties can be achieved through controlled deformation or thermal treatment. Recent work has focused on identifying process parameters, for example, scanning width in additive manufacturing \cite{Popova2017} or the initial texture that can be cold rolled to achieve a desired final texture \cite{jung2019modelling}, but the problem of sequence identification is largely unexplored.  A typical process for manufacturing a component such as a turbine blade can contain as many as twenty distinct processing stages, each taking the microstructure from one state to another, eventually achieving a desired microstructure. The problem of ‘processing path design’ aims to discover a sequence of known processes to realize microstructures with optimal properties \cite{mssl38,mssl60,li2005texture,li2007processing,brough2017microstructure,shaffer2010building,khosravani2017development,yabansu2017extraction}. In \cite{li2007processing}, a model capable of predicting texture evolution starting from any specific deformation path (as processing path lines) was developed to guide process design. Because of non-uniqueness in processing path solution (different processing paths leading to similar microstructural features) and complex nature of the microstructure–property–process relationships, the problem of identifying an optimal process pathline is better solved using data-mining strategies \cite{jung2019modelling, shaffer2010building,mssl5,mssl11}. In \cite{mssl38,mssl60}, this problem was addressed by using proper orthogonal decomposition to project textures obtained from different processes in a reduced order `process plane' where the process sequence needed to obtain a given ODF was obtained via linear programming. In \cite{shaffer2010building}, a large database of process lines was developed and superimposed in a property closure to identify optimal process sequences using graph search techniques.  While these techniques have allowed exploration of innovative process design techniques, they do not yet use the advantages offered by modern machine learning methods for achieving significant model reduction directly in the texture space. For example, Ref. \cite{shaffer2010building} indicates the need for a 12-dimensional subspace to adequately capture the texture dependence of the macroscale plastic properties of interest in fcc metals. 

In recent years, neural network based generative models have become popular and are capable of generating a continuous low dimensional space of input data that can be graphically visualized. In this paper, a Variational Autoencoder (VAE) network is used to achieve a continuous latent space for the entire database. Using proximity analysis in this latent space, multiple process solutions to the problem of texture or property design is demonstrated. To train this network, we focus on the development of a computationally generated open source database of textures of single phase FCC metals that result from sequences of up to four stages of tension, compression and rolling along different directions in various permutations. Such a database can be a platform for collaborative testing and assessing machine learning algorithms by computer scientists who do not have domain knowledge in texture evolution. Visualization of these large databases has been a challenge. In the past, reduced methods such as proper orthogonal decomposition have been used to reduce the texture space \cite{mssl5,mssl11},  but they have usually resulted in large dimensionality of reduced spaces. One task that is accomplished in this work is the direct visualization of this database containing 346200 textures in two dimensions using the autoencoding technique. 


\section{Methodology}
\subsection{Process model used in the database}


Imposed deformation is described by the deviatoric\footnote{Since plastic deformation via dislocation slip is isochoric} velocity gradient tensor, $\mathbf{L}$. The velocity gradient tensor itself can be expressed as a linear combination of multiple processes, as shown in the equation below. The first term in the equation represents tension in the X direction, second term represents rolling along Y (with Z being the short transverse direction), the next three terms represent pure shear along the three directions and the last three terms represent rotations about the three axes. A linear combination of these  terms can be used to describe all possible volume preserving deformations. Tension, compression and rolling being the deformation modes employed in this study, the $\alpha_{i}'$s used to describe these processes are shown in table \ref{processes}.

\begin{table}[ht!]
\begin{minipage}{0.5\textwidth}
\begin{eqnarray}
\textbf{L} \ &=& \  \alpha_{1} \left[ \begin
{array}{ccc}   1 & 0 & 0 \\
           0 & -0.5 & 0 \\
           0 & 0 & -0.5 \end{array} \right] \ \nonumber 
+ \ \alpha_{2}\left[ \begin
{array}{ccc}   0 & 0 & 0 \\
           0 & 1 & 0 \\
           0 & 0 & -1 \end{array} \right] \\
 &+&          
           \alpha_{3}\left[ \begin
{array}{ccc}   0 & 1 & 0 \\
           1 &  0 & 0 \\
           0 & 0 & 0 \end{array} \right] \ \nonumber +
 \ \alpha_{4} \left[ \begin
{array}{ccc}   0 &  0 & 1\\
            0 &  0 & 0 \\
           1 & 0 & 0 \end{array} \right] \\
           &+& \
           \alpha_{5}\left[ \begin
{array}{ccc}   0 & 0 & 0 \\
           0 &  0 & 1 \\
           0 & 1 & 0 \end{array} \right] \ \nonumber
+ \ 
           \alpha_{6}\left[ \begin
{array}{ccc}   0 & -1 & 0 \\
           1 &  0 & 0 \\
           0 & 0 & 0 \end{array} \right] \\
           &+&  \ \alpha_{7}\left[ \begin
{array}{ccc}   0 &  0 & -1\\
            0 &  0 & 0 \\
           1 & 0 & 0 \end{array} \right] \ \nonumber + \
           \alpha_{8}\left[ \begin
{array}{ccc}   0 & 0 & 0 \\
           0 &  0 & -1 \\
           0 & 1 & 0 \end{array} \right]
\label{eq:Matrix1}
\end{eqnarray}
\end{minipage}
\hfill
\begin{minipage}{0.45\textwidth}
\centering
\begin{tabular}{|c|c|}
\toprule
Process   & $\alpha_1,\alpha_2$ \\ 
\midrule
Tension X & 1,0           \\ 
Tension Y & -0.5,0.75      \\ 
Tension Z & -0.5,-0.75   \\ 
Rolling YX    & 1,-0.5        \\ 
Rolling ZY    & 0,1        \\ 
Rolling XZ    & -1,-0.5       \\ 
\bottomrule
\end{tabular}
\caption{Coefficients for describing tension and Rolling in different planes}
\label{processes}
\end{minipage}
\end{table}

Besides  the 6 processes, denoted by $P_{i}$ with $i=[1,6]$ referring to table \ref{processes}, strain rate is another parameter which can be used, and this is denoted by $\beta$. This will control the amount of effective deformation at the end of a time step for any particular process. The set of $\beta$ values chosen in this work are $\beta_i = \{ -1, -0.5, 0.5, 1 \}$. The rolling ZY notation, with a positive $\beta$ denotes that the rolling direction is Y and the compression direction is Z, with no strain in the transverse X direction. On the other hand, rolling ZY notation, with a negative $\beta$ denotes the opposite, i.e., rolling direction is Z and the compression direction is Y.

\subsection{Material model and ODF evolution}
We employ the axis-angle parametrization of the orientation space proposed by
Rodrigues \cite{mssl5} as $\textbf{r} = \textbf{n} \ \text{tan}
(\frac{\theta}{2})$. Given the
Rodrigues vector $\textbf{r}$, the rotation tensor $\textbf{R}$ transforming quantities from the crystal frame to the sample frame is takes the form
\begin{eqnarray}\label{rotation}
&&\textbf{R} =
\frac{1}{1+\textbf{r}.\textbf{r}}(\textbf{I}(1-\textbf{r}.\textbf{r}) + 2(\textbf{r} \otimes \textbf{r} + \textbf{I} \times
\textbf{r})).
\end{eqnarray}
The database contains the fundamental region  of the orientation
space in which each crystal orientation is represented uniquely. The fundamental region for the cubic symmetry group as used in this paper
results in a truncated cube. The ODF value (represented by $\mathcal{A}(\mathbf{r},t)$) defined on the fundamental region describes the probability density corresponding to orientation $\mathbf{r}$ at time $t$. 

The evolution of ODF is governed by the ODF conservation equation similar to the continuity equation in continuum mechanics, which expressed in Eulerian form 
is given by \cite{kumar2000computational}:
\begin{eqnarray}
\label{eq:evolA}
&&\frac{\partial {\mathcal{A}} ({\textbf{r}},
t)}{\partial t} + \bigtriangledown {\mathcal{A}} ({\textbf{r}}, t)
\cdot {{\textbf{v}}} ({\textbf{r}}, t) + {\mathcal{A}} ({\textbf{r}}, t) \bigtriangledown \cdot
\textbf{v}({\textbf{r}}, t) \ = \ 0
\end{eqnarray}
where ${{\textbf{v}}} ({\textbf{r}}, t)$ is the Eulerian
reorientation velocity.

From Eq. \ref{eq:evolA}, it is seen that the
evolution of the ODF  is controlled by the
reorientation velocity ${{\textbf{v}}} ({\textbf{r}}, t)$. The velocity
gradient for the process, $\mathbf{L}$ is linked to ${{\textbf{v}}}
({\textbf{r}}, t)$. The reorientation velocity is evaluated through crystal
constitutive relations, which involve the crystal velocity
gradient. The velocity gradient of a crystal with orientation,
$\textbf{r}$, yields the following form:
\begin{equation}
\myb{L} \ = \ \myb{\Omega} \ + \ \myb{R} \sum_{\alpha}
\dot{\gamma}^{\alpha} \bar{\myb{T}}^{\alpha} {\myb{R}}^{T}
\end{equation}
where $\myb{\Omega}$ is the lattice spin, $\dot{\gamma}^{\alpha}$
is the shearing rate along the slip system $\alpha$ and
$\bar{\myb{T}}^{\alpha}$ is the Schmid tensor for the slip system
$\alpha$, given by $\bigb{ \bar{\myb{m}}^{\alpha} \ \otimes \
\bar{\myb{n}}^{\alpha} }$, where $\bar{\myb{m}}^{\alpha}$ is the
slip direction and $\bar{\myb{n}}^{\alpha}$ is the slip plane
normal, both in the crystal lattice frame. The expressions for the
spin and symmetric parts are obtained as shown below:
\begin{equation}
\myb{\Omega} \ = \ \myb{W} \ - \ \sum_{\alpha}
\dot{\gamma}^{\alpha} \myb{R} \bar{\myb{Q}}^{\alpha} \myb{R}^{T}
\label{eq:skew}
\end{equation}
\begin{equation}
\bar{\myb{D}} \ = \ \sum_{\alpha} \dot{\gamma}^{\alpha} \
\bar{\myb{P}}^{\alpha} \label{eq:Vel_gradsym}
\end{equation}
where $\bar{\myb{P}}^{\alpha}$ and $\bar{\myb{Q}}^{\alpha}$ are
the symmetric and skew parts of the Schmid tensor respectively and
$\bar{\myb{D}}$ is the macroscopic deformation rate expressed in
the lattice frame through, $\bar{\myb{D}} \ = \ \myb{R}^{T}
\myb{D} \myb{R}$. The shearing rate on slip systems is given by a
power law  and we further assume that all slip systems have
identical hardness.
\begin{equation}
\dot{\gamma}^{\alpha} \ = \ \dot{\gamma}^{0} \left({
\frac{\tau^{\alpha}}{s}} \right)^\frac{1}{m} sgn
\left(\bigb{\frac{\tau^{\alpha}}{s}} \right ) \label{eq:PowerLaw}
\end{equation}
where $s$ is the slip system hardness, $m$ is the strain rate
sensitivity, $\dot{\gamma}^{0}$ is a reference rate of shearing
and ${\tau}^{\alpha}$ is the resolved shear stress on slip system
$\alpha$. Further, the resolved stress is related to the crystal
Cauchy stress as
\begin{equation}
\tau^{\alpha} \ = \ \bar{\myb{\sigma}} \dotp
\bar{\myb{P}}^{\alpha} \label{eq:cauchy}
\end{equation}
By solving the system of equations (\ref{eq:Vel_gradsym} $-$
\ref{eq:cauchy}), the crystal cauchy stress ($\bar{\myb{\sigma}}$)
and the shear rate ($\dot{\gamma}^{\alpha}$) can be evaluated.
Next, using Equation (\ref{eq:skew}), we can evaluate the lattice
spin vector as,
\begin{equation}
\omega \ = \ \mbox{vect}\bigb{\myb{\Omega}}
\end{equation}
which is then used to evaluate the reorientation velocity as,
\begin{equation}
\textbf{v} \ = \ \frac{1}{2} \bigb{\omega \ + \ (\omega \cdot
\textbf{r})\textbf{r} \ + \ \omega \otimes \textbf{r}}
\label{eq:reorient_velocity}
\end{equation}

Finally, the ODF, $\mathcal{A}$, over the current fundamental
region ${\mathcal{R}}$ is evaluated from the Eulerian form
(Equation (\ref{eq:evolA})) of the conservation equation. 
Eq. (\ref{eq:evolA}) is solved using a finite element formulation
developed in ref. \cite{kumar2000computational}.

\subsection{Database structure}
\label{database_structure}
\begin{figure}[!ht]
  \centering
  \includegraphics[width=0.7\textwidth]{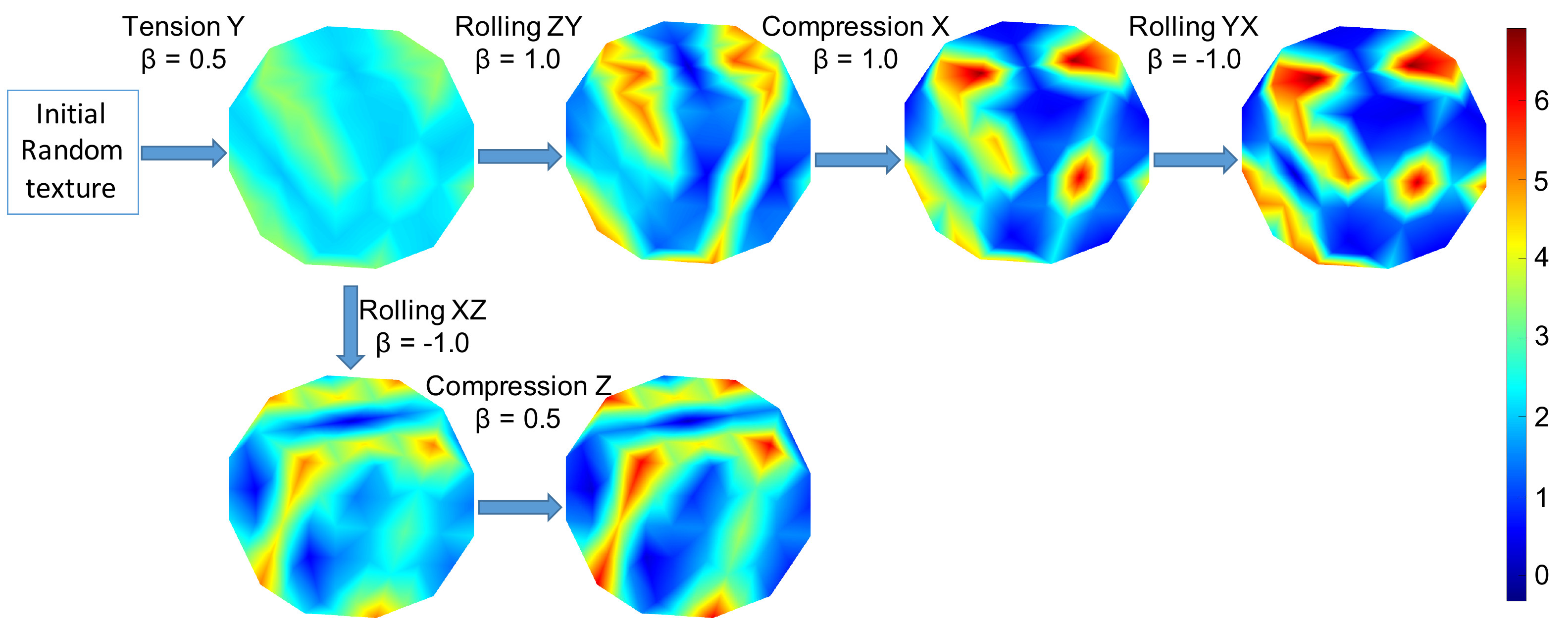}
  \caption{ODF evolution starting from an initial random texture through two process sequences. Both sequences share the same first stage, with the top depicting a four stage sequence and the bottom row depicting three stage sequence.}
  \label{example_sequence}
\end{figure}

With a combination of four strain rates and six possible processes, there are a total of 24 processes that can be used in all permutations in each stage of a process sequence to make up the process sequence-ODF database. In the current work sequences containing up to 4 stages have been considered. An example of what the transformation of ODF after each stage in a sequence of three stages can be seen in figure \ref{example_sequence}. This gave a total of 346,200 sequence-ODF pairs and this was split into 250,000 data points for the training and a little less than 50,000 data points for the validation and test sets.

\subsection{Variational auto encoder}
An auto encoder is a learning model which consist of an encoder and a decoder neural network. The encoder takes the input data from a high dimensional space and maps it to a code which is generally of lower dimension. The decoder takes this lower dimensional code and maps it back to the original space. Training of the model is performed by reducing the error between what is input to the encoder and what is output by the decoder. While a one-one mapping is a possible outcome, this is avoided by having a code dimension smaller than the input dimension. The objective is to let the model uncover useful properties in the input data while the  decoder output only serves the purpose of finding the encodings. When the encoder and decoder networks are single layer perceptron with linear activation functions and a mean squared loss function, the model is reduced to a principal component analysis (PCA)\cite{autoencoder,Goodfellow-et-al-2016}.

\begin{figure}[!ht]
  \centering
  \includegraphics[width=0.7\textwidth]{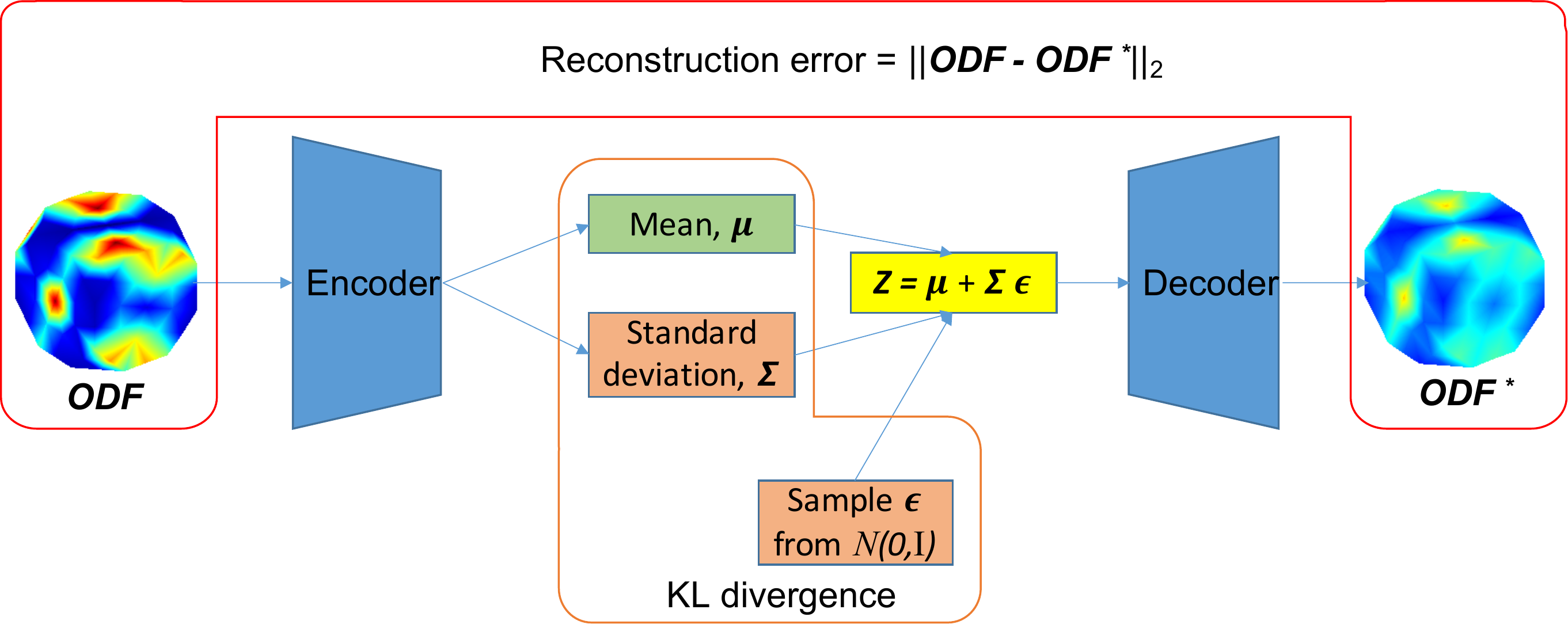}
  \caption{Schematic of a VAE showing reconstruction of an ODF and error measures used for training the model}
  \label{vaeschematic}
\end{figure}

A variational auto encoder (VAE) is a class of auto encoders where the latent representation is space spanning, and can be used as a generative model \cite{kingma2013autoencoding,doersch2016tutorial}. Figure \ref{vaeschematic} shows a schematic of a VAE. The model is trained by minimizing both the reconstruction error and a Kullback–Leibler (KL) divergence. This divergence measures how the mean and standard deviation from the encoder diverges from the target distribution from which the input data is hypothesised to be generated. A latent space thus ``learnt" can be used in interpolatory and extrapolatory modes and hence the model's generative capabilities. VAEs have been employed in many domains: chemical design by generating new drug-like molecules and optimizing for properties \cite{automaticchemical}, designing new materials possessing desired optical absorption properties \cite{Stein2018}, and discovering hypothetical metastable vanadium oxides \cite{Noh2019}. 

Thus, VAEs give a reduced order representation of the data set which can be visualized on low dimensional plots and any point can be sampled from the low dimensional space to produce an output in the original space. This makes VAE attractive to use on the database described in section \ref{database_structure}, for dimensionality reduction, reconstruction and proximity analysis based on common features.

\begin{figure}[h]
\centering

\subfloat[]{%
  \includegraphics[width=0.46\linewidth]{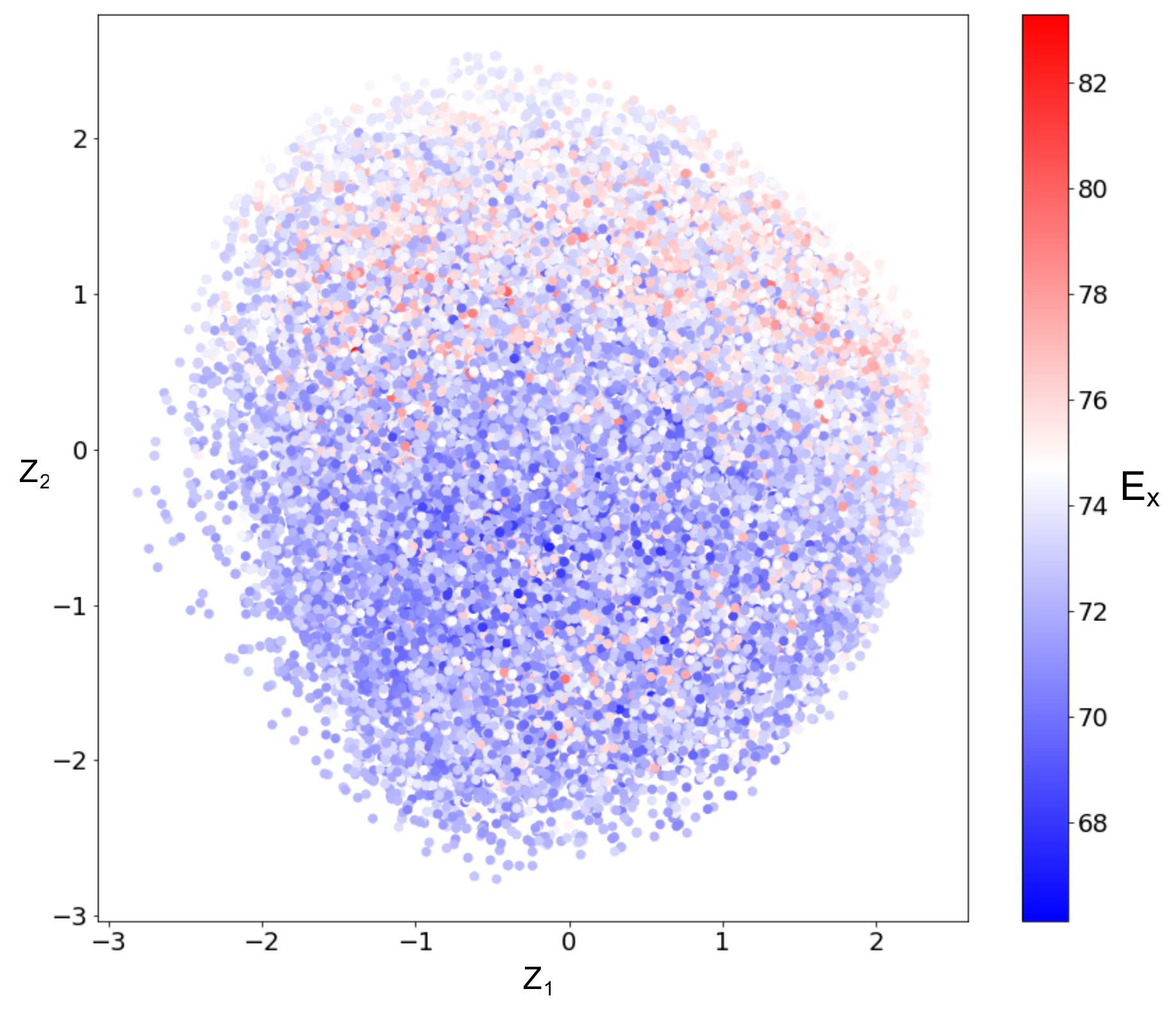}%
  \label{latentspace_Ex}%
}\qquad\hspace{0.2cm}
\subfloat[]{%
  \includegraphics[width=0.46\linewidth]{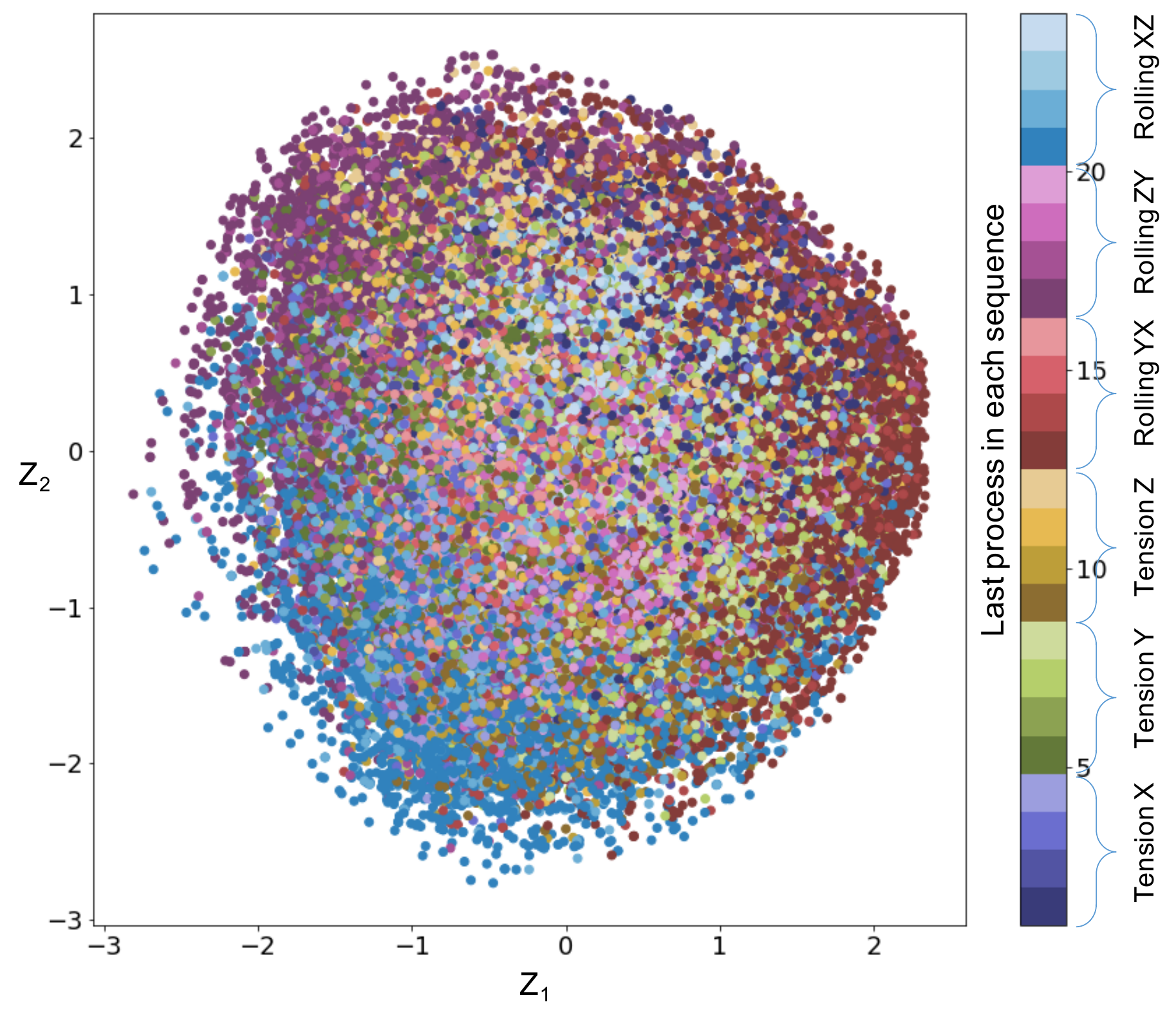}%
  \label{latentspace_p1}%
}
\caption{Each point in the latent space representation corresponds to one ODF from the training set. Figure (a) shows points colored by E$_x$ and figure (b) shows points colored by the last process from each sequence. }
\label{latentspace_training}
\end{figure}

For each ODF input to the model, a vector containing the latent space coordinates (same as the mean in figure \ref{vaeschematic}) is obtained when parsed through the encoder. In the case of a 2D latent space this can be visualized on a 2D scatter plot and figure \ref{latentspace_training} shows the ODFs from the training set mapped onto the latent space. The points are colored by Young's modulus along the X direction in figure \ref{latentspace_Ex}, and it is seen that there are certain regions that contain hot spots. Figure \ref{latentspace_p1} shows the points colored by the last process from the sequence which resulted in that particular texture. The supplementary video shows how specific data points are clustered together based on the last process of the sequence.

\begin{figure}[H]
\centering

\subfloat[]{%
  \includegraphics[width=0.50\linewidth]{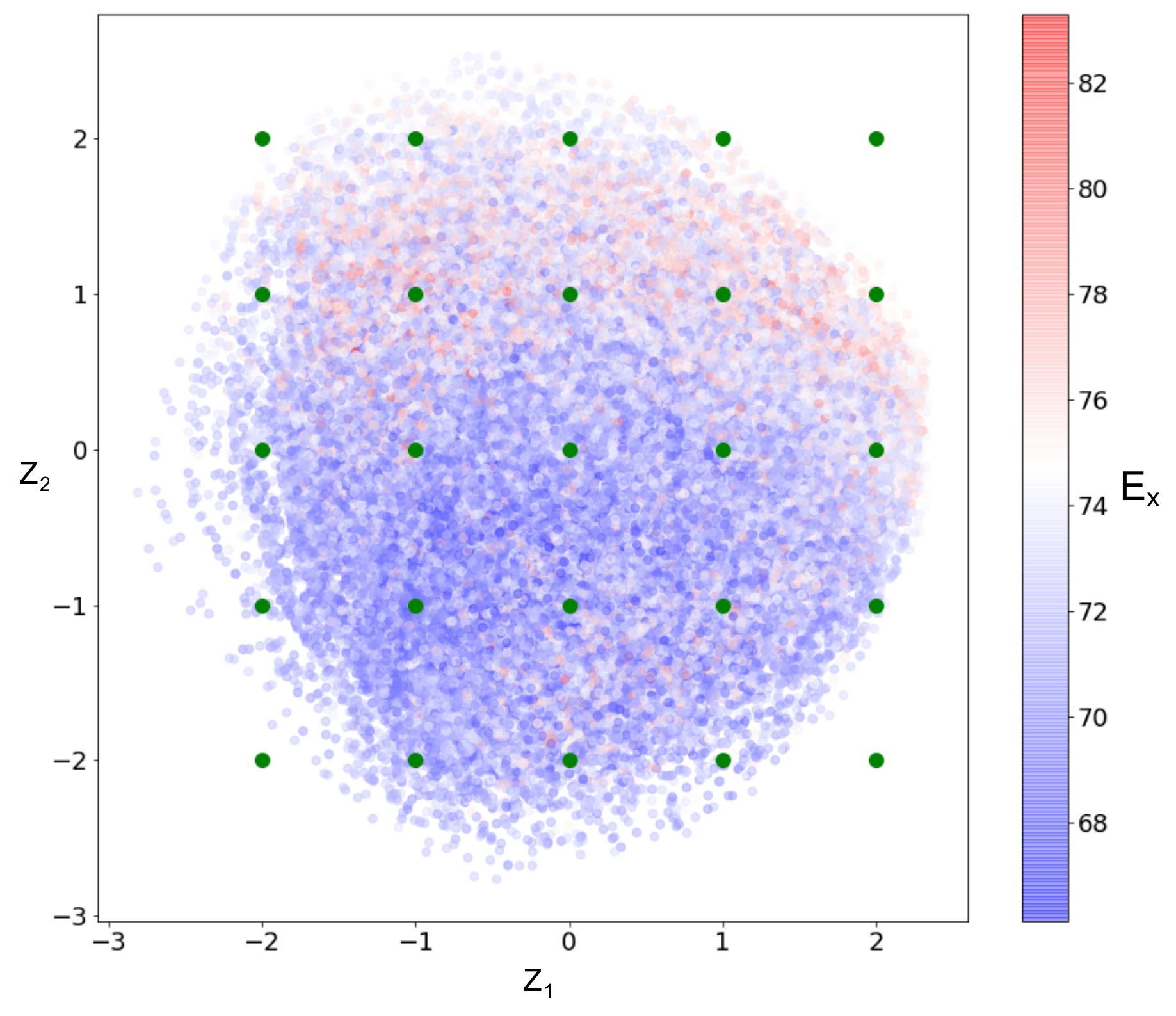}%
  \label{generationpoints}%
}
\hspace{0.03\linewidth}
\subfloat[]{%
  \includegraphics[width=0.43\linewidth]{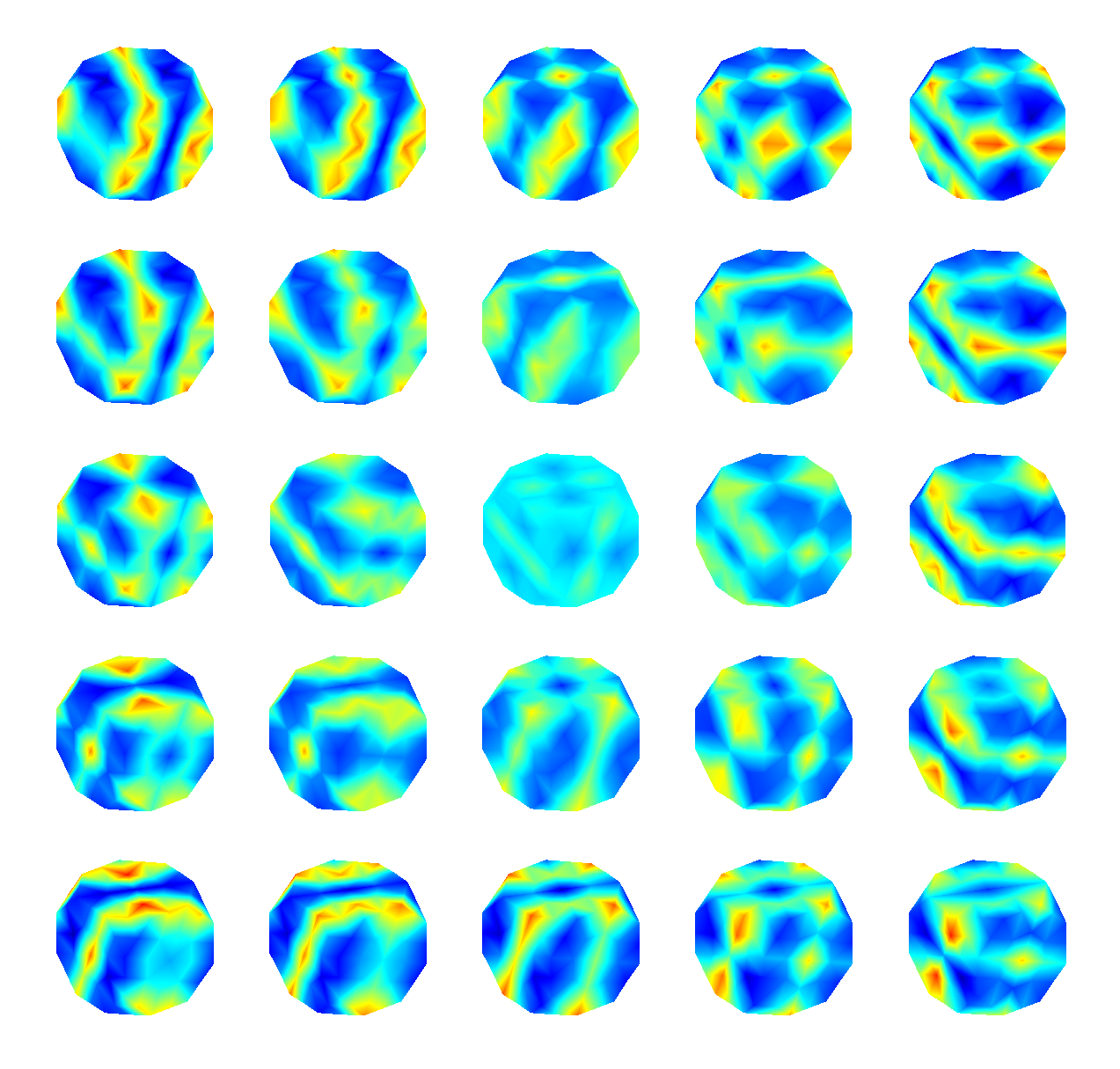}%
  \label{generationODFs}%
}
\caption{This figure illustrates the generative capabilities of the trained model, including extrapolatory reconstruction. Figure (a) depicts the points from the 2D latent space that were sampled for reconstruction and figure (b) shows reconstructed ODFs.}
\label{generativecapability}
\end{figure}

The model being generative, points can be sampled from a grid on the latent space and parsed through the decoder, to understand the spatial dependence of ODF mapped on the latent space. Figure~\ref{generativecapability} shows this sampling. It is seen that the corners represent different families of material textures, and farther away from origin, higher the intensity at particular nodes in the ODF.
Different regions in the latent space are seen to encode specific features in the ODF space, and this feature separation will be useful for analysis of performance as shown in the next section. 

\section{Results and discussion}

\begin{figure}
\centering

\subfloat[]{%
  \includegraphics[width=0.36\linewidth]{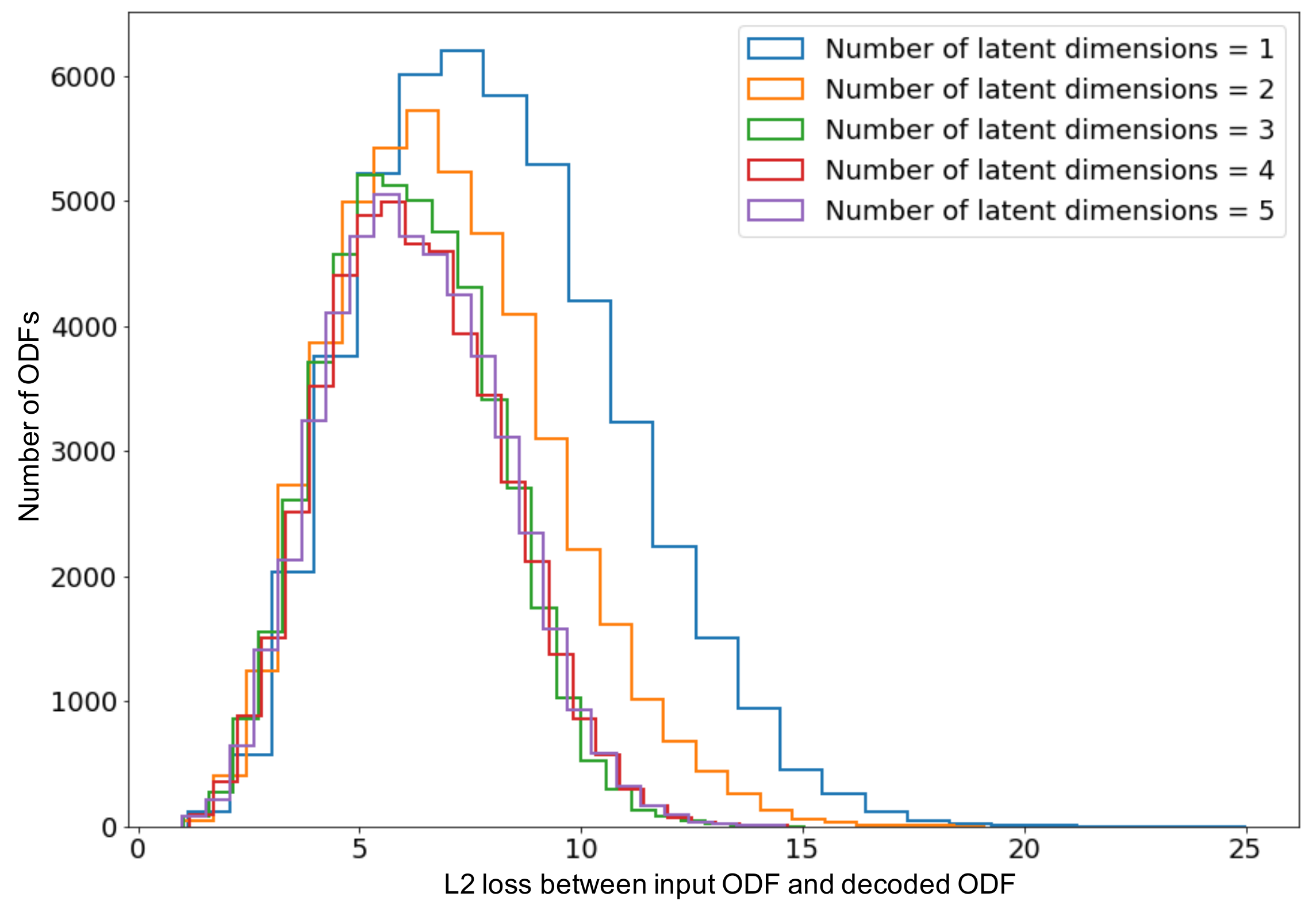}%
  \label{odfl2error}%
}
\hspace{0.03\linewidth}
\subfloat[]{%
  \includegraphics[width=0.55\linewidth]{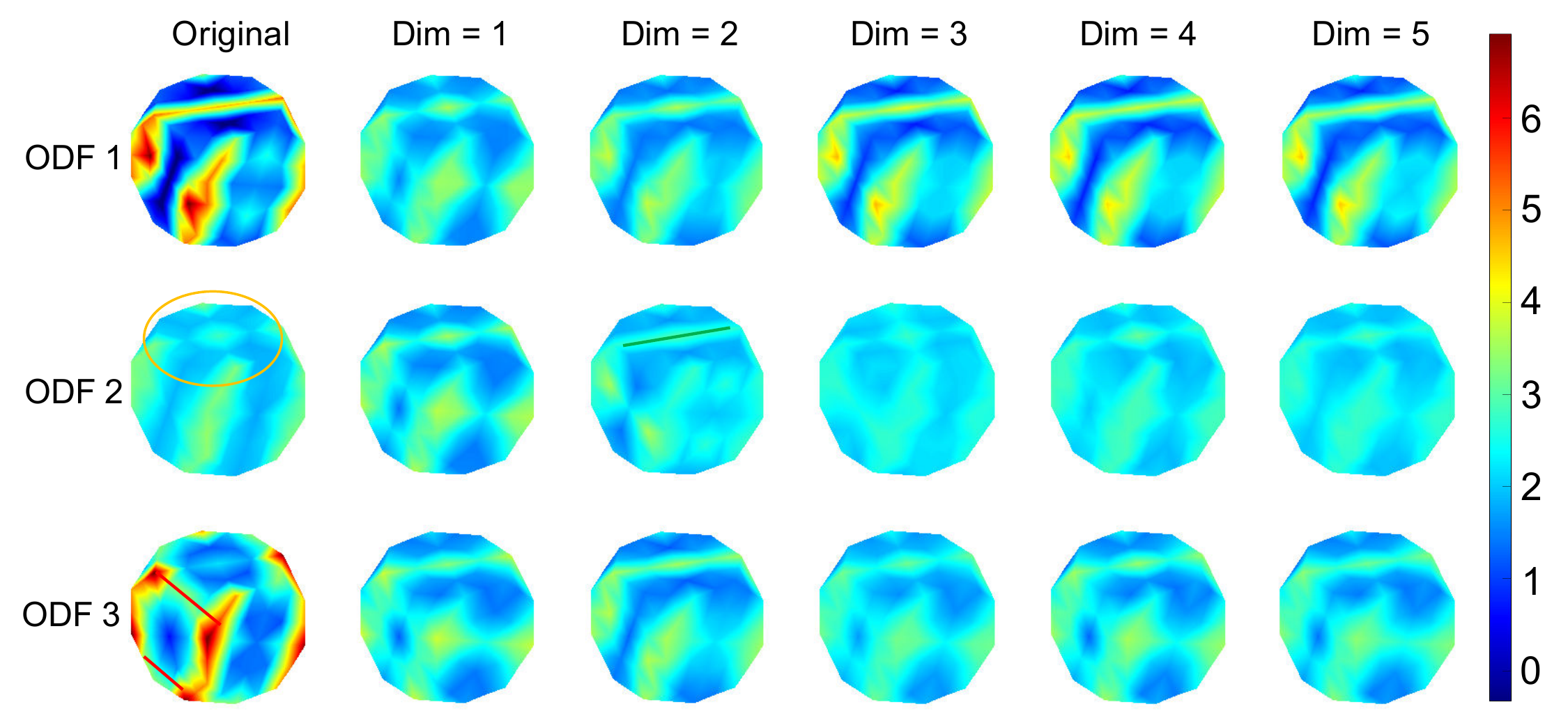}%
  \label{odfreconstruction}%
}
\caption{Quantitative and qualitative study of VAE performance to understand optimal number of dimensions in the latent space. Figure (a) shows different histograms of L2 error between ODFs in the validation dataset and their reconstruction for varying latent dimensions. Figure (b) shows 3 ODFs to illustrate preservation of important features, marked with fiber lines, by the encoder and subsequent reconstruction.}
\label{latentdimensionchoice}
\end{figure}

One hyperparameter that needs to be identified is optimal dimension of the latent space. In this vein, multiple VAEs were trained with the latent space ranging between 1 and 5 dimensions. One of the main criteria for a ``well trained" model for the current application would be the optimal identification of important features from the dataset. For this purpose, the quantitative estimate through distribution of reconstruction error and qualitative analysis by way of capturing important ODF features are carried from the different trained VAEs. The validation dataset is used for this purpose. Figure \ref{odfl2error} captures the variation of reconstruction error histogram, vector norm of the difference between input ODF and its reconstruction, with dimension of the latent space. It is seen that there is a loss in intensity at many nodes in the ODF. This is attributed to the scaling of input data, and descaling of the output. This is required to keep all attributes of the input data between 0 and 1, which is one of the assumptions of the methods used, and the NN layers. 

Since capturing useful features of data into the latent space is more important than exact reconstruction, figure \ref{odfreconstruction} shows a study of model performance by looking for the preservation of important fibers from Rodrigues space. For this purpose three ODFs are used as inputs to different models, and lines shown on these are the important fibers whose preservation is an important metric of model performance. From ODF 1, it is seen that all the models are able to capture the most important features, but with varying intensity matching. From ODF 2, dim = 2 captures the top layer marked as a single fibre whereas the other models are able to replicate the input ODF better. In ODF 3, looking at the red lines, it is again not sufficiently reproduced in dim = 2 and it gets better with increasing dimensions. Though dim = 1 is able to capture many features, it does not sufficiently show diversity in the patterns. 

Together, figure \ref{latentdimensionchoice} shows that the most optimal choice of latent space dimension is 3, considering the trade-off between size and performance. A 2D latent space performs nearly close to a 3D latent space in most of the cases. Since this paper aims at understanding the database through a graphical treatment, and the spatial separation in the latent space, the results going further are shown from a 2D latent space, with one example of prediction from a 3D latent space.  

\begin{figure}
  \centering
  \includegraphics[width=0.95\textwidth]{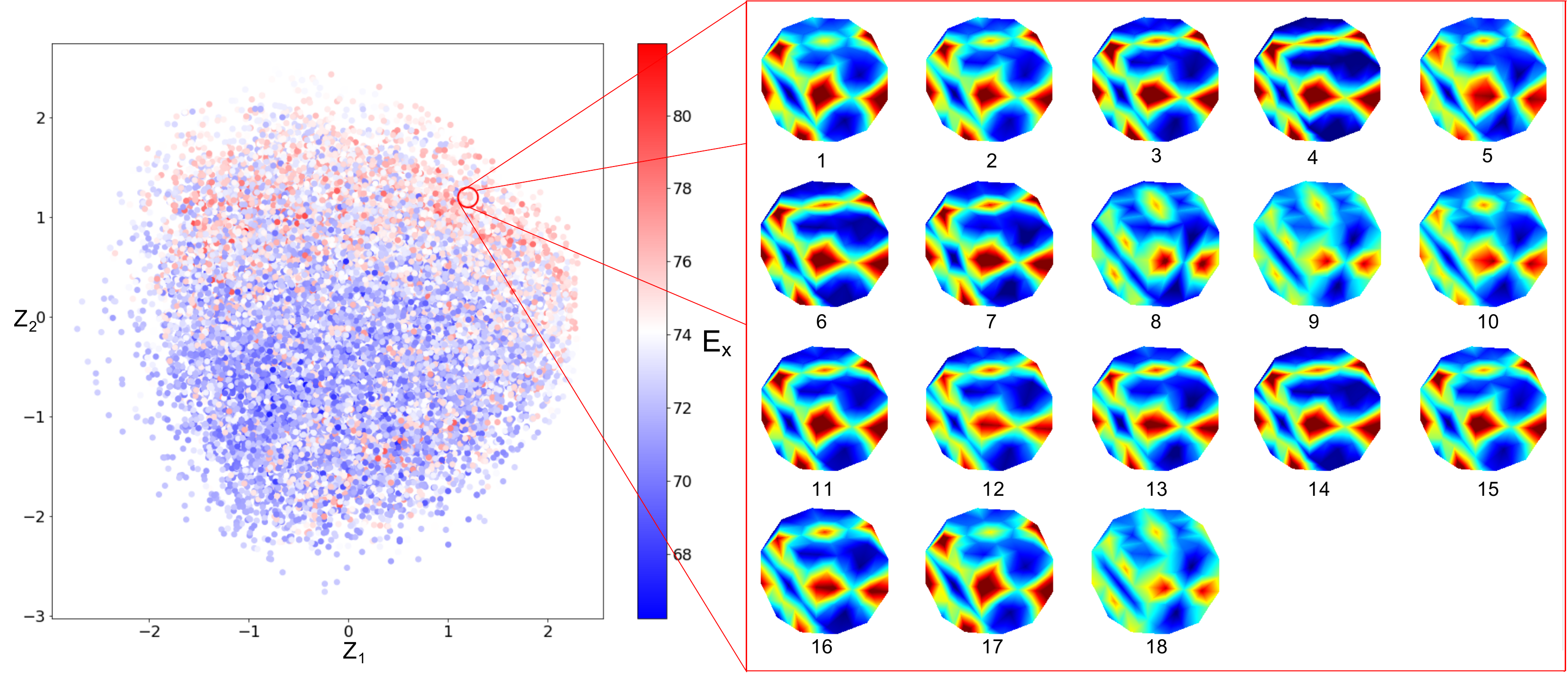}
  \caption{ODFs from the validation set encoded onto latent space. The red circle marks a circular area with radius 0.1 units around the point (1.2,1.2) in the latent space and ODFs encoded in that region are shown on the right.}
  \label{windowODFs}
\end{figure}

The approach to finding a process sequence for an unknown ODF involves mapping this ODF onto the latent space and identifying other ODFs in the vicinity. In that vein, it would be necessary to understand the similarity of ODFs in a small window in the latent space, and the processes which resulted in those particular structures. 
Figure~\ref{windowODFs} shows the ODFs in one such circular window with radius 0.1 units. It is seen that all ODFs identified in the area, exhibit very similar features and fall into two different sets of similar ODFs. 
Table \ref{windowprocesses} shows the processes that gave rise to these textures and the effective elastic modulus in x direction (E$_x$). The properties are obtained as a polycrystal average by an integral over the fundamental region: $\bar{\myb{C}} = \int \limits_{{\mathcal{R}}} \myb{C} \ {\mathcal{A}} ({\textbf{r}}) \ dv$. The stiffness constants for FCC Aluminum are used here. The elastic modulus is computed as $\myb{E} \ = \ \frac{1.0}{{\bigb{\bar{\myb{C}}}}^{-1}_{(11)}}$. Even though the ODFs exhibit very similar features, their E$_x$ values show a significant spread. This can be attributed to small variations at certain nodes which arise from distinct processes and their order as well. Thus, using this strategy, it is possible to identify ODFs and sequences which are similar, but also identify process sequences and resultant textures which give rise to better properties.

\begin{table}
\centering
\begin{tabular}{|c|c|c|c|c|c|c|c|c|c|}
\toprule
{} & Stage 1 &  $\beta_{1}$ & Stage 2 &  $\beta_{2}$ & Stage 3 &  $\beta_{3}$ & Stage 4 &  $\beta_{4}$ & E$_x$\\
\midrule
1  &  Rolling YX &    -1.0 &   Tension Z &     1.0 &  Rolling YX &    -0.5 &   Tension Z &     0.5 &  73.197 \\
2  &   Tension Z &     0.5 &   Compression X &    0.5 &   Tension Z &     1.0 &  Rolling YX &    -1.0 &  73.977 \\
3  &   Tension Z &     0.5 &   Tension Z &     1.0 &  Rolling YX &    -1.0 &   Compression X &    1.0 &  75.368 \\
4  &   Compression X &    1.0 &  Rolling YX &    -1.0 &   Compression X &    11.0 &   Tension Z &     1.0 &  76.811 \\
5  &  Rolling XZ &     1.0 &   Compression Y &     1.0 &  Rolling YX &    -1.0 &  Rolling ZY &    -1.0 &  73.295 \\
6  &   Compression X &    1.0 &   Tension X &    0.5 &  Rolling YX &    -1.0 &   Tension Z &     1.0 &  75.908 \\
7  &  Rolling YX &    -1.0 &  Rolling ZY &    -1.0 &  Rolling YX &    -1.0 &  Rolling XZ &     1.0 &  75.543 \\
8  &   Compression X &    1.0 &  Rolling YX &    -1.0 &  Rolling XZ &    -0.5 &  Rolling ZY &    -1.0 &  74.941 \\
9  &  Rolling ZY &    -0.5 &  Rolling ZY &    -0.5 &  Rolling YX &    -0.5 &  Rolling YX &    -0.5 &  73.925 \\
10 &   Compression X &    0.5 &   Tension Z &     0.5 &  Rolling YX &    -1.0 &  Rolling ZY &    -0.5 &  74.404 \\
11 &   Tension Z &     1.0 &   Compression X &    1.0 &  Rolling YX &    -1.0 &   Tension Z &     0.5 &  74.956 \\
12 &   Compression Y &    0.5 &   Tension Z &     1.0 &   Compression X &    1.0 &  Rolling YX &    -1.0 &  76.001 \\
13 &  Rolling ZY &    -1.0 &  Rolling YX &    -1.0 &  Rolling XZ &     1.0 &  Rolling YX &    -0.5 &  75.259 \\
14 &   Compression X &    1.0 &  Rolling YX &    -1.0 &   Compression X &    0.5 &   Tension Z &     1.0 &  75.909 \\
15 &   Tension Z &     0.5 &   Tension Z &     1.0 &  Rolling YX &    -1.0 &   Compression X &    0.5 &  74.108 \\
16 &  Rolling YX &    -1.0 &  Rolling ZY &    -0.5 &   Compression X &    1.0 &   Tension Z &     0.5 &  75.400 \\
17 &   Tension Z &     0.5 &  Rolling YX &    -0.5 &  Rolling YX &    -1.0 &   Tension Z &     1.0 &  73.229 \\
18 &   Compression Z &    0.5 &  Rolling ZY &    -1.0 &   Tension Z &     0.5 &  Rolling YX &    -1.0 &  74.395 \\
\bottomrule
\end{tabular}
\caption{Process sequences leading to ODFs in the red area from figure \ref{windowODFs}.}
\label{windowprocesses}

\end{table}

\begin{figure}
\centering

\subfloat[]{%
  \includegraphics[width=0.54\linewidth]{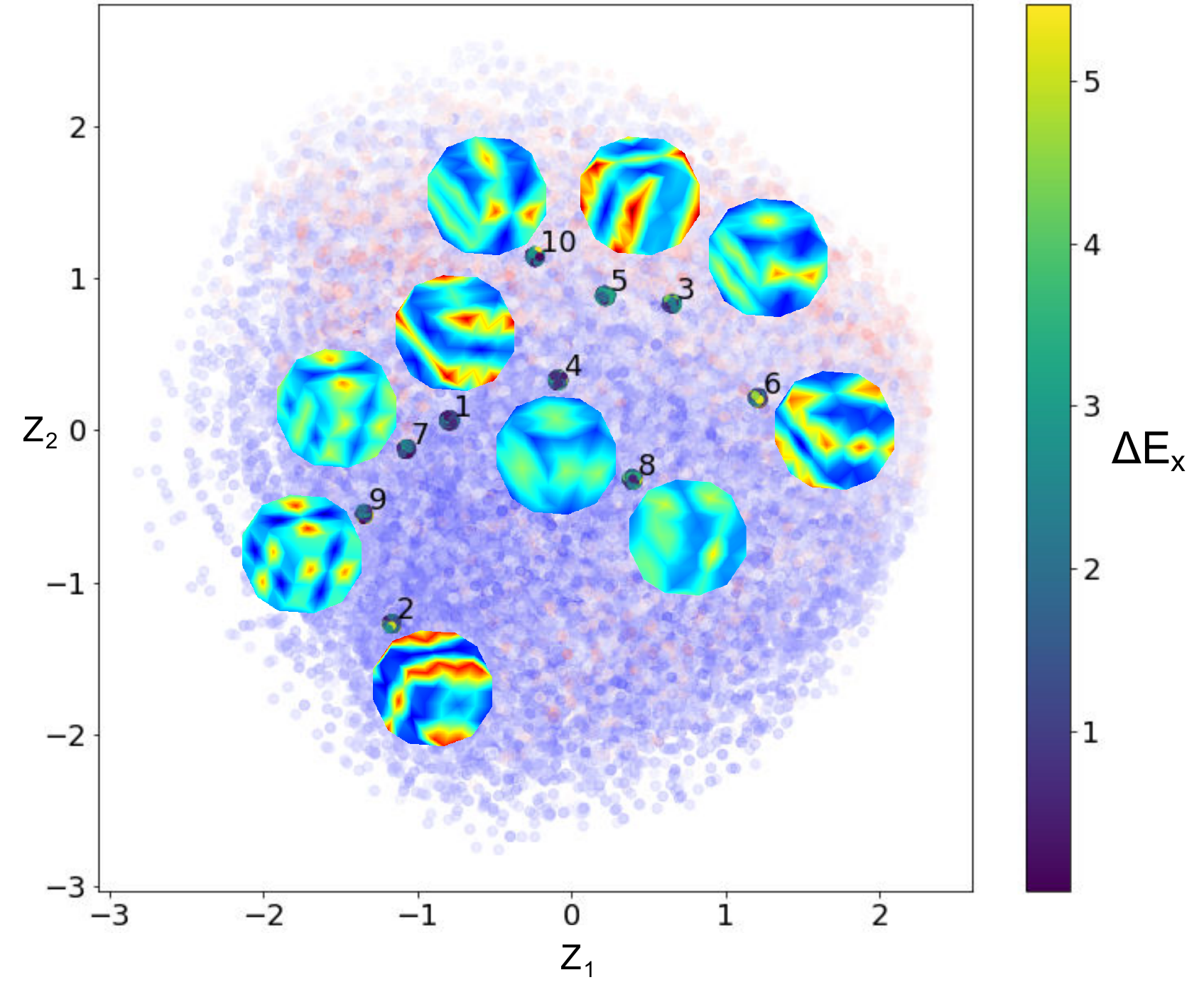}%
  \label{2Dinputandlatentspace}%
}
\hspace{0.03\linewidth}
\subfloat[]{%
  \includegraphics[width=0.42\linewidth]{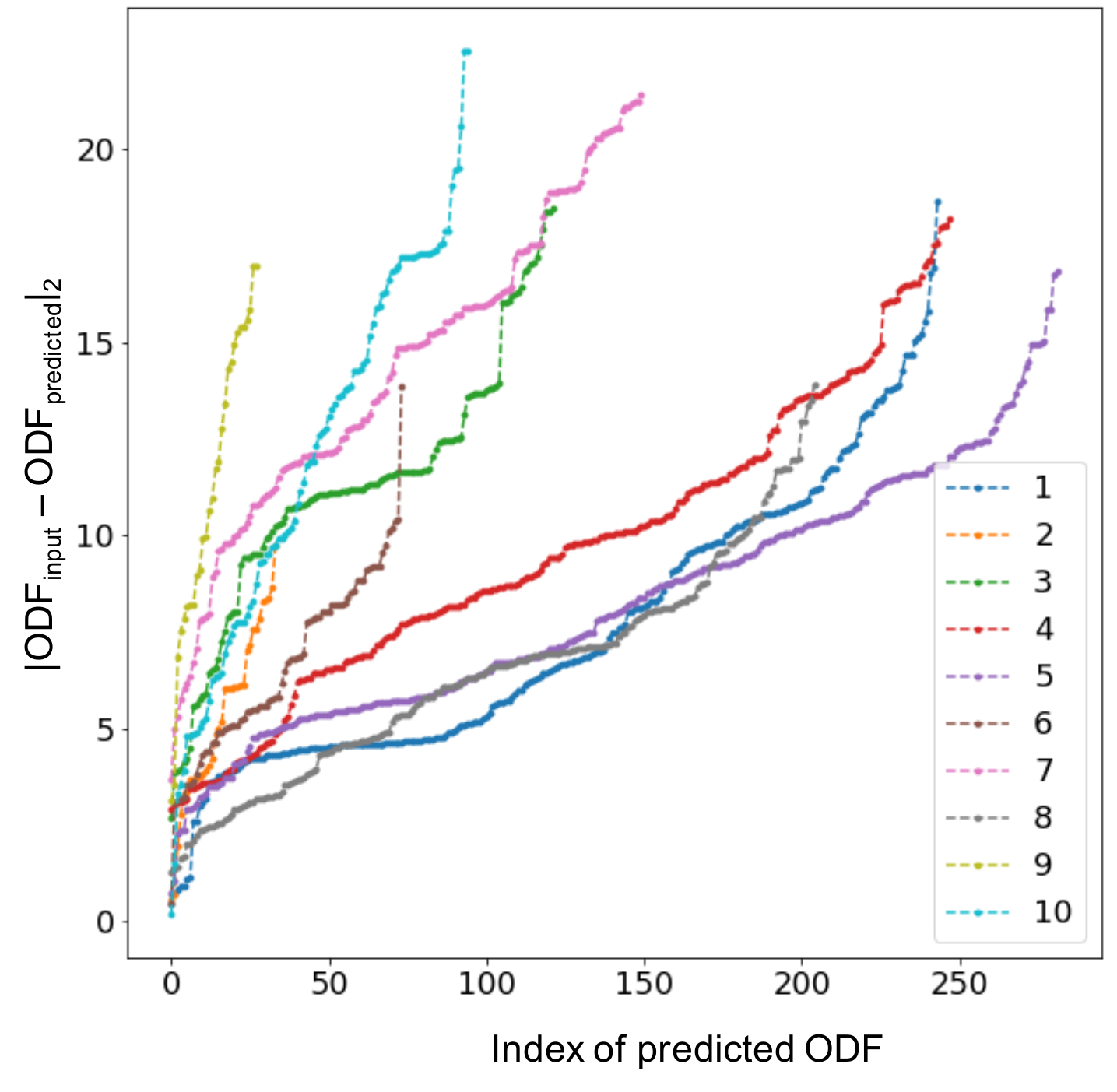}%
  \label{2Dpredicterror}%
}
\caption{Figure (a) shows a few ODFs used for sequence identification from the test set. The encodings from these are overlayed on top of encodings from training set, and points from the training set in a small area around the test encodings (predicted ODFs) are colored by the difference in E$_x$ from test ODFs. Figure (b) shows the L2 difference between input and predicted ODFs for each test case.}
\label{2Dpredictlatentspaceanderror}
\end{figure}

Figure \ref{2Dinputandlatentspace} shows ten ODFs from the test set mapped onto the latent space. The region shows latent coordinates of ODFs from the training set. Next to each ODF, there is a cluster of points which represent ODFs in close proximity to the input ODF, as described previously. These points are colored by the difference between E$_{x}$ of input ODF and ODFs to which they correspond. For simplicity these ODFs will be called `predicted ODFs' going forward. Figure \ref{2Dpredicterror} shows L2 error between each input ODF and its corresponding predicted ODFs. The number of predicted ODFs depends on the density of points in the latent space, and since the density closer to origin is higher, test cases 1,4, and 8 have the highest number of predicted ODFs.

\begin{figure}
  \centering
  \includegraphics[width=0.95\textwidth]{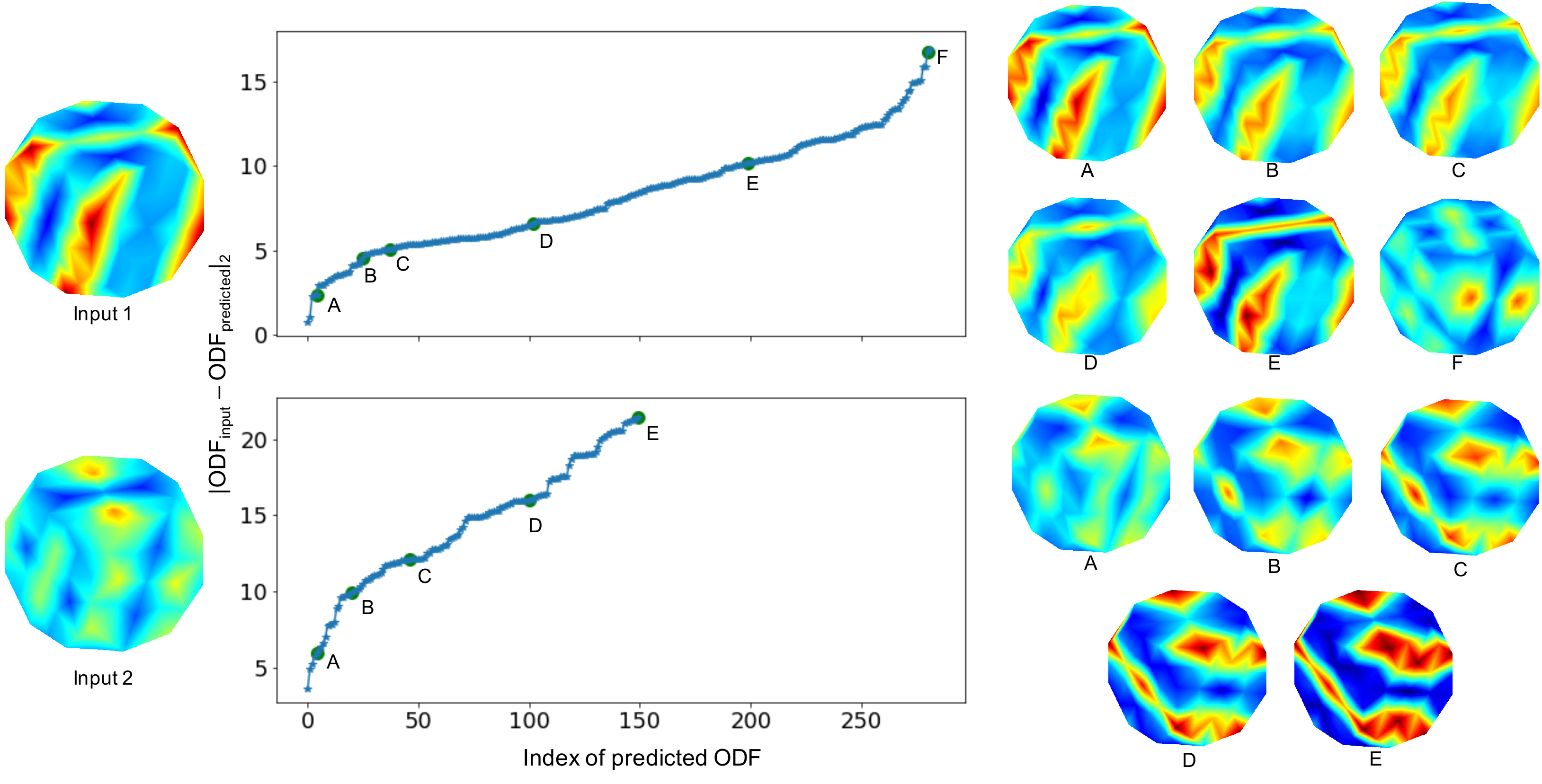}
  \caption{Two cases to illustrate the sequence prediction task through ODF identification. ODFs on the right are a few matched points from the training set, which have varying L2 errors with respect to the input, indicated by numbers on the error curve.}
  \label{2Dpredict_withidentifiedODFs}
\end{figure}

To illustrate sequence identification, the predicted ODFs for inputs 5 and 7 from figure \ref{2Dpredictlatentspaceanderror} are examined. These will be referred to as inputs 1 and 2 going forward. Figure \ref{2Dpredict_withidentifiedODFs} shows the two inputs, along with 6 and 5 predicted ODFs for inputs 1 and 2 respectively. The L2 difference between input and each predicted ODF are highlighted by green dots in both cases. Even though a few ODFs are exactly similar to the input, the L2 difference  is not zero because of small perturbations. Tables \ref{predictinputssequences}, \ref{processes2Dcase1}, and \ref{processes2Dcase2} show the process paths that led to these ODFs. The tables show that the model is capable of identifying ODFs which have resulted from sequences with lesser stages for the illustrated cases. While 1A, 1B, and 1C are all identically similar to the input, 1D, 1E, and 1F are not. The latter three have been identified because they preserve the important features of input, though they have different intensities. 1B, 1C, and 1D have only 3 stages in their respective processes, while the input has 4 stages.

\begin{table}
\centering
\begin{tabular}{|c|c|c|c|c|c|c|c|c|}
\toprule
{} & Stage 1 &  $\beta_{1}$ & Stage 2 &  $\beta_{2}$ & Stage 3 &  $\beta_{3}$ & Stage 4 &  $\beta_{4}$ \\
\midrule
1 &  Tension Z &     1.0 &   Compression X &    0.5 &   Compression Z &    1.0 &  Rolling XZ &    -0.5 \\
2 &  Tension X &     1.0 &  Rolling ZY &    -0.5 &  Rolling XZ &     0.5 &   Compression Y &    1.0 \\
\bottomrule
\end{tabular}
\caption{Process sequences for the input cases from figure \ref{2Dpredict_withidentifiedODFs}}
\label{predictinputssequences}
\end{table}

\begin{table}
\centering

\begin{tabular}{|c|c|c|c|c|c|c|c|c|}
\toprule
{} & Stage 1 &  $\beta_{1}$ & Stage 2 &  $\beta_{2}$ & Stage 3 &  $\beta_{3}$ & Stage 4 &  $\beta_{4}$ \\
\midrule
A &   Tension Y &     0.5 &   Tension Z &     1.0 &   Tension X &     0.5 &  Rolling XZ &     1.0 \\
B &  Rolling XZ &     1.0 &   Tension Y &    -1.0 &   Tension Y &     1.0 &   Tension Z &     0.5 \\
C &  Rolling ZY &     0.5 &  Rolling XZ &     1.0 &   Tension Z &     1.0 &         NaN &     NaN \\
D &   Tension Y &     0.5 &  Rolling XZ &     1.0 &  Rolling ZY &    -0.5 &         NaN &     NaN \\
E &  Rolling XZ &     0.5 &   Tension X &    -1.0 &   Tension Y &    -0.5 &         NaN &     NaN \\
F &   Tension X &    -1.0 &   Tension Y &     0.5 &  Rolling ZY &    -1.0 &  Rolling XZ &    -0.5 \\
\bottomrule
\end{tabular}
\caption{Process sequences predicted for the first test case from figure \ref{2Dpredict_withidentifiedODFs}}
\label{processes2Dcase1}

\end{table}

\begin{table}
\centering

\begin{tabular}{|c|c|c|c|c|c|c|c|c|}
\toprule
{} & Stage 1 &  $\beta_{1}$ & Stage 2 &  $\beta_{2}$ & Stage 3 &  $\beta_{3}$ & Stage 4 &  $\beta_{4}$ \\
\midrule
A &   Compression X &    0.5 &  Rolling ZY &    -1.0 &   Tension X &     1.0 &   Compression Z &    0.5 \\
B &  Rolling YX &     1.0 &  Rolling ZY &    -0.5 &  Rolling XZ &    -0.5 &         NaN &     NaN \\
C &   Tension Z &     1.0 &   Compression Z &    1.0 &   Tension X &     1.0 &   Compression Y &    1.0 \\
D &   Tension X &     0.5 &   Compression Y &    1.0 &  Rolling YX &     0.5 &   Tension X &     0.5 \\
E &  Rolling XZ &    -1.0 &   Compression Y &    1.0 &  Rolling YX &     1.0 &  Rolling YX &     1.0 \\
\bottomrule
\end{tabular}
\caption{Process sequences predicted for the second test case from figure \ref{2Dpredict_withidentifiedODFs}}
\label{processes2Dcase2}
\end{table}

\begin{figure}
\centering
\includegraphics[width=0.65\linewidth]{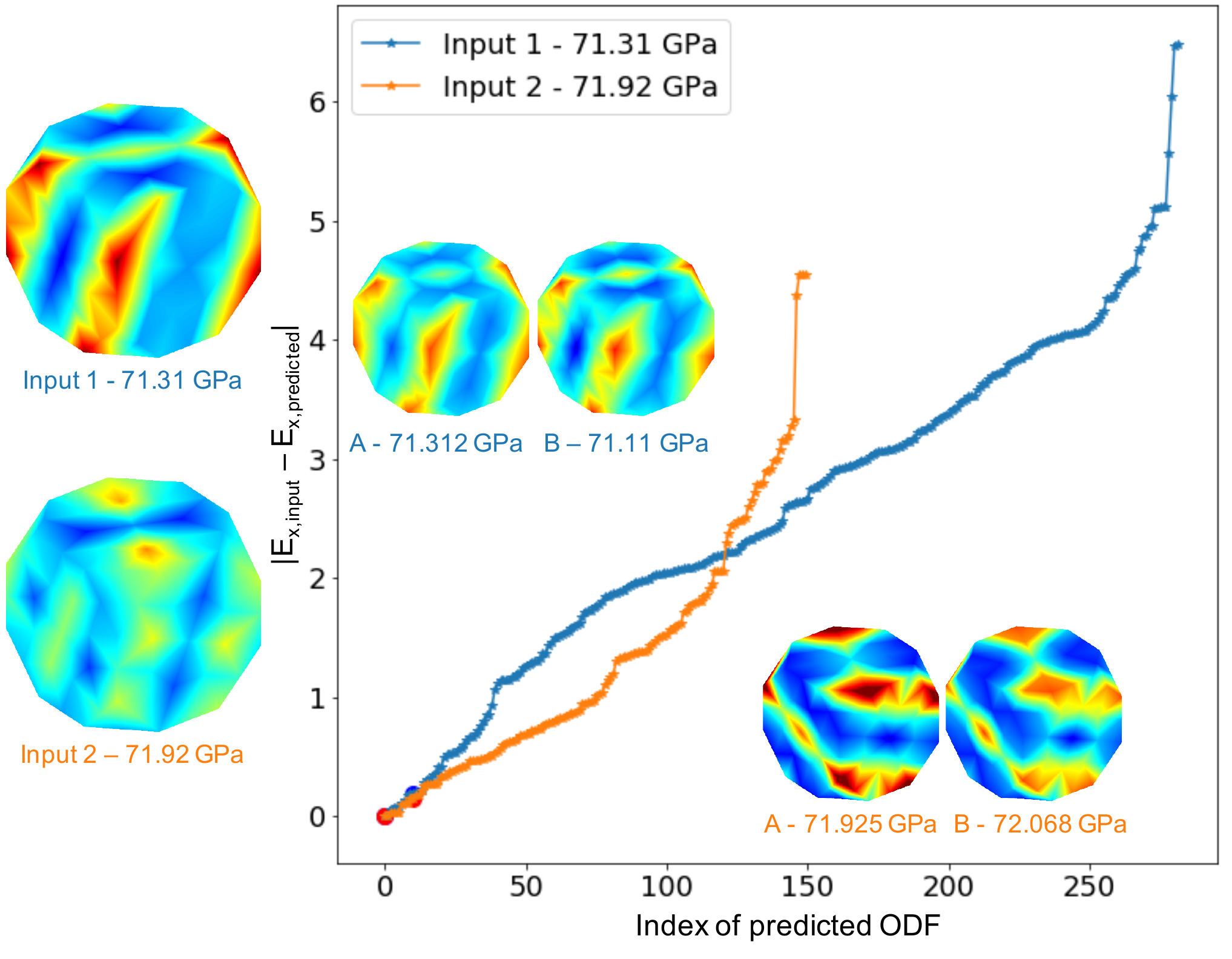}%
\caption{ODF identification by matching E$_{x}$ between input and prediction}
\label{exxplot}%
\end{figure}

The database and trained model can also be used for material design by finding new ODFs in the neighborhood with approximately same property of interest. Figure \ref{exxplot} shows the same two inputs from the previous case, but the plot shows difference between E$_{x}$ of input ODFs and their respective predicted ODFs. Also shown are two predicted ODFs for each case whose E$_{x}$ is very close to their respective inputs, and these points correspond to the blue and red dots on respective curves. It is interesting to see that for both the cases, the ODFs shown are not similar to the inputs. There is either a difference in the features or the intensity. But, the difference in E$_{x}$ is less than 0.5 GPa. The graphs also show that there can be multiple ODFs from the $|\Delta$ E$_{x}|<$ 1 GPa range to choose for predicted ODFs.

\begin{figure}
  \centering
  \includegraphics[width=0.95\textwidth]{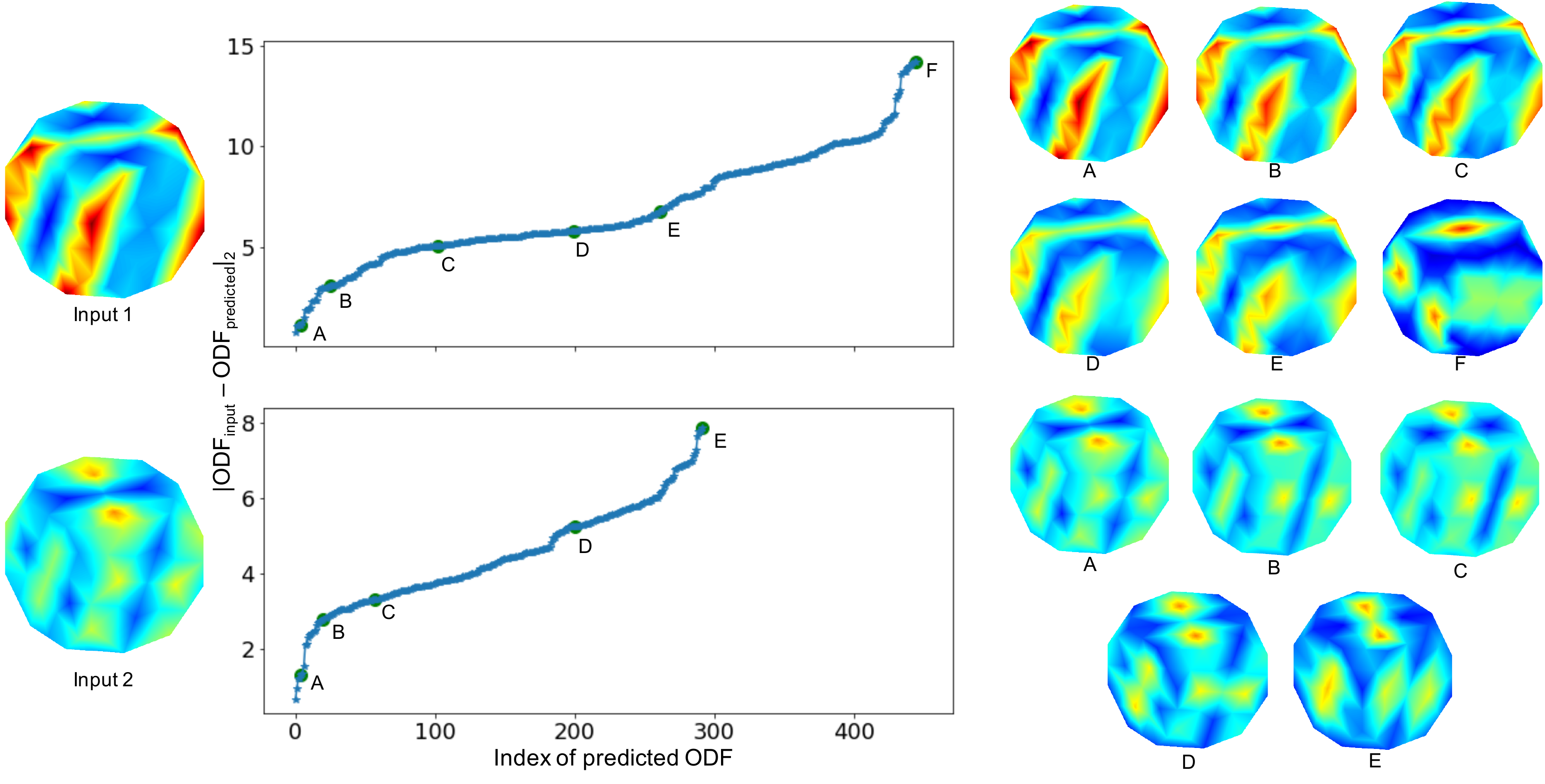}
  \caption{ODF identification task from a 3D latent space}
  \label{3DODFidentification}
\end{figure}

For visualizing the distribution of ODFs and for illustrating proximity, using a 2D latent space is more useful. But, from the latent dimension comparison exercise, a 3D latent space was identified to be most optimal for reconstruction and feature capture. Thus, the prediction task was carried out using a model with a 3D latent space as shown in figure \ref{3DODFidentification}. Similar to the circular area used for the previous case, here the ODFs within a spherical volume with radius 0.2 units from the input ODF were selected for prediction. While the cut-off is arbitrarily chosen, number of predicted ODFs for both inputs is higher. Despite this, the maximum L2 error is lesser compared to a 2D latent space for both inputs.

\section{Conclusions}
The database conceived and disseminated here is a first step in the efforts to unearth PSP linkages. A few features and uses of this database has been explored here. A visual treatment of the database was presented using a VAE for dimensionality reduction through a latent space. The latent space representation exhibits a spatial distribution of ODF based on important features. Moreover, nearly identical ODFs are mapped to the same locations in this space. By searching in the vicinity of an ODF mapped in the latent space for other similar ODFs, it has been shown that it is possible to identify processing routes. A few of these processing routes were also shown to produce similar ODFs in using fewer process stages. Further work will involve training the VAE to use process sequences as an input to the model. This will then be used to identify new process sequences for ODFs identified from the latent space through interpolatory and extrapolatory modes. In the field of natural language processing, sequence to sequence prediction (seq2seq) is a technique used to predict a new word in a sentence, given the previous words \cite{sutskever2014sequence}. This technique can be used to iteratively predict each stage in a sequence and this will also be explored in the future. 

\section{Data Availability}
The raw data and processing scripts required to reproduce these findings are available to download from \url{https://github.com/sriharisundar/processtexturedatabase}.

\section*{Acknowledgements}
The authors would like to acknowledge the Air Force Office of Scientific Research Materials for Extreme Environments Program (Grant No. FA9550-18-1-0091) for financial support. Additionally, the computations have been carried out as part of research supported by the U.S. Department of Energy, Office of Basic Energy Sciences, Division of Materials Sciences and Engineering (Award no. DE-SC0008637) that funds the PRedictive Integrated Structural Materials Science (PRISMS) Center at the University of Michigan. SS would like to thank the open source knowledge contributors which non-exhaustively includes software developers, blog writers, and those who answer questions on online forums.

\bibliographystyle{elsarticle-num}
\bibliography{references}

\end{document}